\documentclass[preprint,12pt]{elsarticle}

\usepackage{amssymb}
\usepackage{amsmath}

\journal{Nuclear Instruments and Methods in Physics Research A}

\begin{document}

\begin{frontmatter}



\title{
Design and implementation of the constant fraction discriminator for glass MRPC timing
}


\author[3]{L.L.~Kurchaninov}
\author[1]{E.A.~Ladygin}
\author[1]{V.P.~Ladygin}
\author[2]{A.A.~Semak}

\affiliation[1]{organization={Joint Institute for Nuclear Research},
            city={Dubna},
            postcode={141980}, 
            state={Moscow region},
            country={Russian Federation}}
\affiliation[2]{organization={Institute for High Energy Physics, National Research Center Kurchatov Institute},
            city={Protvino},
            postcode={142281}, 
            state={Moscow region},
            country={Russian Federation}}
\affiliation[3]{organization={TRIUMF },
            city={Vancouver},
            postcode={BC V6T 2A3}, 
            country={Canada}}
\begin{abstract}
The analog front-end electronics based on the constant fraction discrimination method
is designed and optimized for the Multigap Resistive Plate Chamber (MRPC) timing measurements.  
The total time resolution of $\sim$40~ps has been obtained for 10 and 12 gaps MRPCs 
using cosmic setup and a muon beam at the IHEP U-70 accelerator in Protvino, which complies with the conditions of the SPD experiment at NICA. 
\end{abstract}
 



\end{frontmatter}



\section{Introduction}
\label{intro}

High energy and heavy ion experiments require good particle identification based on the 
time-of-flight (TOF) techniques. MRPCs are widely used
in large area TOF systems due to its good time resolution, high detection efficiency and 
relatively low cost production. In particular, MRPCs are used for particle identification 
at ALICE \cite{alice1,alice2,alice3}, HARP \cite{harp}, STAR \cite{star}, PHENIX \cite{phenix},
BM@N \cite{kuzmin, BMNspect} and other experiments.

The Spin Physics Detector \cite{SPD_TDR} is a universal facility for studying  the nucleon spin structure and other spin related phenomena with polarized proton and deuteron beams placed at the second interaction point at NICA \cite{nica}. The SPD TOF system  will also be based on  MRPCs.
The purpose of the MRPC system is to make a 3$\sigma$ separation of $\pi$/$K$ and $K$/$p$   in the momentum range up to  few GeV/$c$. The required time resolution of the SPD   MRPC system is  $\sim$60 ps \cite{SPD_TDR}.
 
Typically, the  total time resolution of  MRPC  systems 
achieved in present experiments  is 50$\div$70 ps. The best time resolution of $\sim$20 ps 
has been obtained for an MRPC having 24 gas gaps
with a width of 160 $\mu$m  built of thin 400 $\mu$m "soda-lime" glass sheets after the correction  for the time slewing  \cite{best_TI}.
The time resolution of an MRPC is  mainly determined  by the 
detector intrinsic resolution, which  depends on the design and physics of the 
gas discharge;  
the jitter of the front-end electronics (FEE) and cables;  and the
time-to-digit converter (TDC) channel uncertainty. 

Improvement of the MRPC time resolution  requires   correction for the time slewing,
which arises from the signal time delay dependence on its amplitude.  
Therefore, the time of the leading edge has to be corrected
for the signal amplitude. 
Nowadays, the MRPC signal amplitude is usually estimated using the 
fast Time-Over-Threshold (ToT)
method,  which only reads out the threshold crossing time and the signal
time interval over the threshold. 
FEE based on  NINO ASIC with an embedded ToT 
function \cite{NINO,usenko}
and  TDC system built using the HPTDC ASIC \cite{alice-DAQ} 
can provide   contribution to the total time resolution of $\sim$20~ps \cite{best_TI}. 
Another way is to analyze the MRPC signal waveform. 
In particular, the NA61 MRPC system \cite{na61test,na61} is based on 
the use of  a fast analog amplifier and   DRS4 module \cite{DRS} 
for the signal waveform analysis. 
The time resolution of the fast analog FEE and 
high-speed DRS4 based \cite{DRS0} waveform digitization module, integrating the waveform
correction and filtering, digital discrimination and  linear interpolation
is better than 10~ps \cite{WDM}.
A new analysis method based on a neural network and machine learning algorithms was
proposed and implemented to make the best use of the MRPC signal waveform \cite{WF0,WF1}.
The MRPC time resolution obtained by this method using 7 points on the signal  leading edge
was found  better than provided by the traditional ToT  \cite{WF1}.

The constant fraction discrimination  method \cite{CFD}
is widely used for TOF measurements \cite{Knoll}.  
The fast discriminator which triggers at a constant fraction of the input signal amplitude allows to obtain the optimum time resolution  regardless
of the signal amplitude. In this case there is no need for 
the time slewing effect correction.
Moreover, a constant fraction discriminator (CFD) is suitable 
for large scale systems due to its simplicity and low cost.
  
This paper presents first results on the design, optimization and first application of 
the FEE based on the constant fraction discrimination method for the SPD MRPC system \cite{SPD_TDR}.  
The paper is
organized as follow. Section \ref{MRPC} describes the MRPC design used for the studies and analysis of their  signal shapes.
The constant fraction discriminator concept and schematics
is discussed in Section \ref{design}. 
The results of the cosmic muons and muon 
beam test at the U70 accelerator  are described
in Section \ref{cosmics} and Section \ref{beamtest}, respectively.  The conclusions are
drawn in the last Secton.

\section{MRPC}
\label{MRPC}
 
\begin{figure}[hbtp] 
\centering
\includegraphics*[width=80mm,height=80mm]{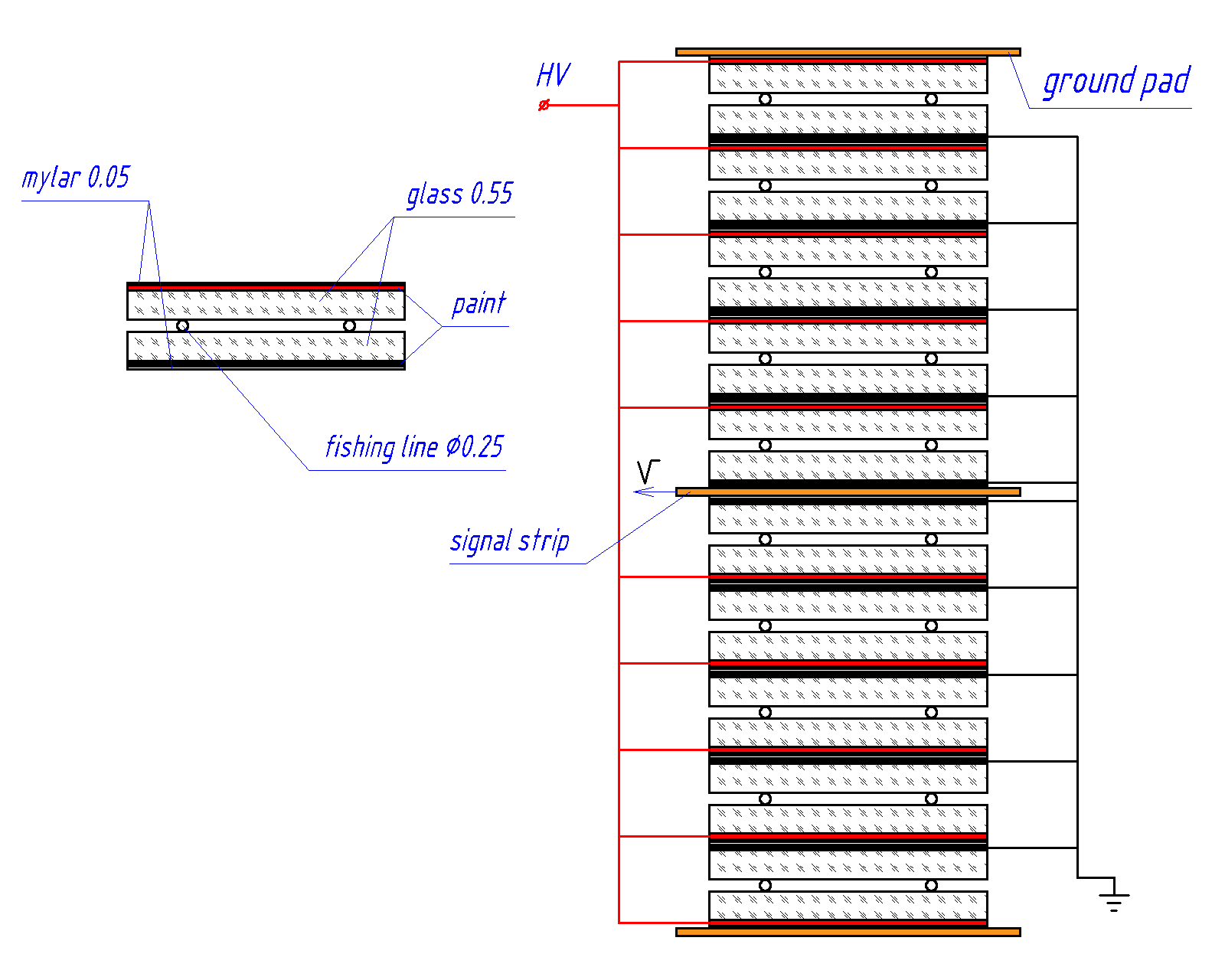}
 \caption{The schematic view of a 10-gap MRPC. }
\label{fig:fig1} 
\end{figure}

Four MRPCs  were manufactured for the present studies. 
The two studied MRPCs have 10  and 12 gas gaps in total, while two others used
for the trigger purpose have 10 gas gaps. 
Each  MRPC consists of two identical 5- or 6-gap stacks with an anode strip readout plate in between.  The schematic  cross section of a 10-gap MRPC is shown in Fig.\ref{fig:fig1}.

Each gas gap is formed by two 0.55~mm thick glass plates with a bulk resistivity of 3$\times$10$^{13}$ $\Omega$cm. The gap between the glass plates
is defined by a spacer made of  fishing line 250~$\mu$m   
in diameter. Graphite conductive coating with a surface 
resistivity  of 2--5~M$\Omega$/square is painted on outer surfaces
of the stacks to distribute high voltage to create an electric field in the sensitive area.
The anode  readout plate is a one-sided   PCB with a thickness of 100~$\mu$m. The thickness of the copper coating is 35~$\mu$m.
The sensitive area of the MRPC is 16$\times$35.1 cm$^2$. The MRPC has 32
10$\times$160 mm$^2$ readout strips with 1~mm gaps between them.
The signals are read on both ends of the anode strips.
Each MRPC is enclosed
in a gas-tight aluminium box. The bottom of the box
is made of a double-sided PCB (motherboard) with a
thickness of 2.5~mm. The top of the box is covered by
an aluminium plate 1.5~mm thick.
The MRPCs were operated at a high voltage across the gap up to 3 kV 
with $\sim$50 cc/min flow of 90\% C$_2$F$_4$H$_2$, 5\% SF$_6$ and 5\% C$_4$H$_{10}$ gas mixture.
 
The simulation based on the simple Townsend model \cite{townsend} of gas discharge  
shows that the intrinsic time resolution of an MRPC with {10 gas gaps of 250 $\mu$m thickness each} is $\sim$20 ps \cite{simulation}.

\begin{figure}[hbtp] 
\centering
\includegraphics*[width=140mm,height=70mm]{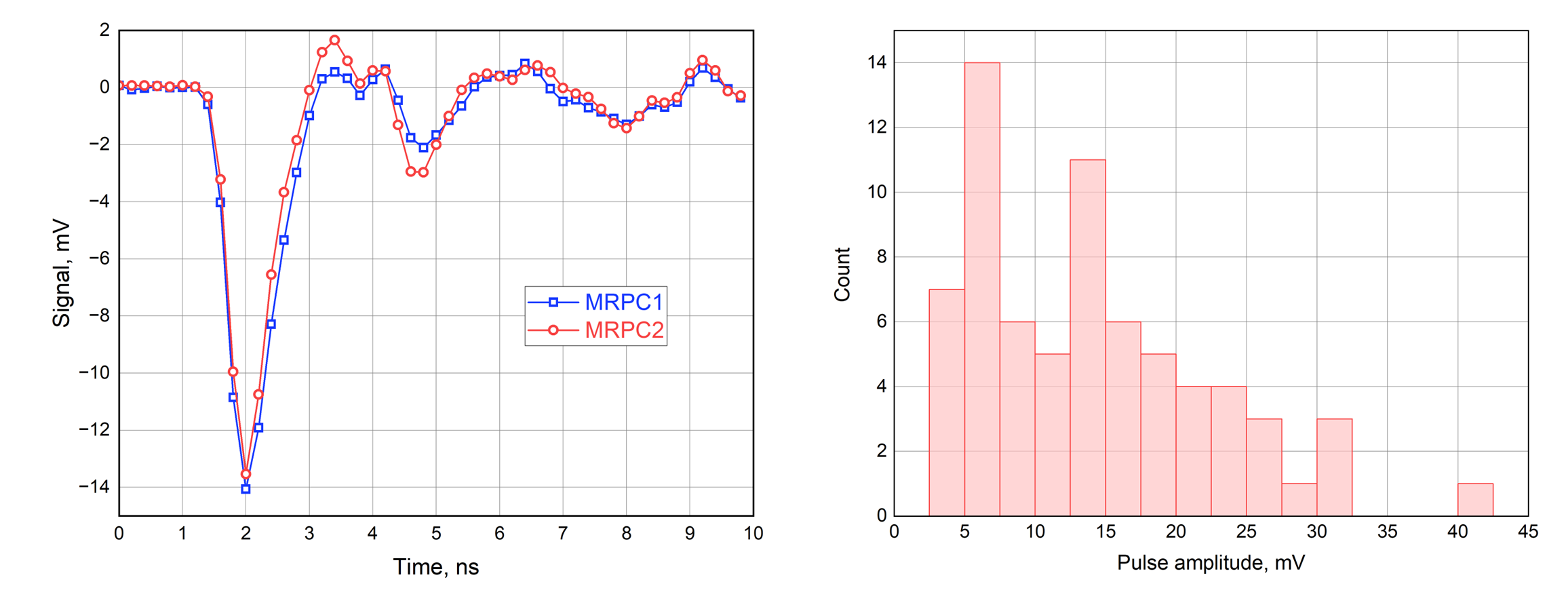}
 \caption{Averaged signals from the 12-gap and 10-gap MRPCs shown by the squares and circles, respectively (left panel). The   signal  amplitudes from the 12-gap MRPC (right panel). }
\label{fig:fig2} 
\end{figure}

The MRPC signal shape was studied with cosmic muons. 
For this purpose, {the two studied and two trigger} MRPCs were stacked vertically 
on the top of each other. 
The MRPCs were operated at a high voltage across the gap of 3 kV.
The coincidences of the signals from the two outer  chambers were used as a trigger for the  Tektronix DPO4104B oscilloscope (bandwidth 1~GHz, 5~GS/sec),  while the waveforms of the signals from both ends of the central strips of the two inner chambers 
were recorded by the LabView program. The results of the measurements for 100 triggers are presented in Fig.\ref{fig:fig2}. The left panel demonstrates the averaged waveforms of the  signals
for the 12-gap (MRPC1) and 10-gap (MRPC2) MRPCs shown by the squares and circles, respectively. 
The observed signal reflections are due to  mismatch of the anode strip line impedance ($\sim$65 $\Omega$) and  50 $\Omega$ termination of the oscilloscope input.
The right panel of Fig.\ref{fig:fig2} represents the distribution of the signal  amplitudes from the 12-gap MRPC. The discrimination threshold was set to 0.9~mV while only signals with amplitude above 2~mV were selected for the analysis.

The waveforms shown in  Fig.\ref{fig:fig2}  were used to reconstruct the MRPC original signal shape. The oscilloscope and MRPC contributions were taken as a single pole and a triangular (linear rise and linear fall) shape, respectively. Best fit gives the rise  and fall  times as  416~ps and 509~ps, respectively,  with a 160~ps time constant for a single pole  corresponding to the oscilloscope bandwidth of 1~GHz.

\section{FEE concept and schematics} 
\label{design}

\begin{figure}[hbtp] 
\centering
\includegraphics*[width=140mm,height=50mm]{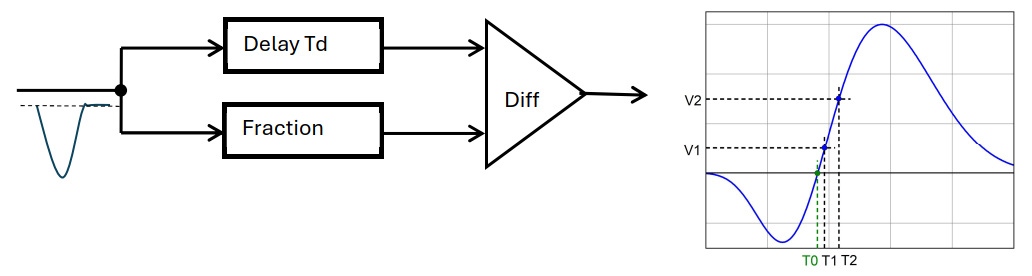}
 \caption{Conceptual diagram of a CFD (left panel) and  zero-crossing  point finding 
for two positive thresholds (right panel). }
\label{fig:fig3} 
\end{figure}

CFDs are widely used for time measurements \cite{CFD,Knoll}. Its conceptual diagram is shown in Fig.\ref{fig:fig3}. The incoming signal is split in two identical copies: one is delayed, and another one is inverted and attenuated by a defined factor. 
Difference of the two signals has zero-crossing point independent on the signal amplitude, thus suppressing the time slewing, present in simple leading-edge discriminators. 
Timing is then determined with a zero-cross circuit which is usually a Schmitt-trigger with the lower threshold set to zero. In present study an alternative method  explained in the right panel of  Fig.\ref{fig:fig3} is used. Instead of zero-crossing point finding, two leading edge comparators  with two thresholds marked in the plot as $V_1$ and $V_2$ are used. The linear combination of the corresponding time points, $T_1$ and $T_2$, is used to reconstruct the signal time reference. 

This combination might be, e.g., an extrapolation to zero threshold emulating a conventional CFD, or it can be optimized for best time resolution. 
One of the thresholds can have a negative value, however,     
the thresholds with both positive values were used in the present design. 
The advantage of this method is a flexibility in optimization of the time resolution, important at the R\&D stage. The obvious disadvantage is the doubling of TDC channels.

The zero-crossing time $T_0$ can be calculated (assuming the linear signal rise and negligible electronics noise) as
\begin{eqnarray}
\label{t0}
T_0 =\frac{V_2\cdot T_1 - V_1\cdot T_2}{V_2 -V_1} = T_1 - Q\cdot | T_2 - T_1 |,  
\end{eqnarray}
where $Q$=$1/(R-1)$ and $R$=$V_2/V_1$>1.  Time $T_0$ depends  linearly on the 
$(T_2-T_1)$ difference.  The case of the nonlinear $T_0$ dependence which provides a better time resolution  is described in Section 5.

\begin{figure}[hbtp] 
\centering
\includegraphics*[width=140mm,height=60mm]{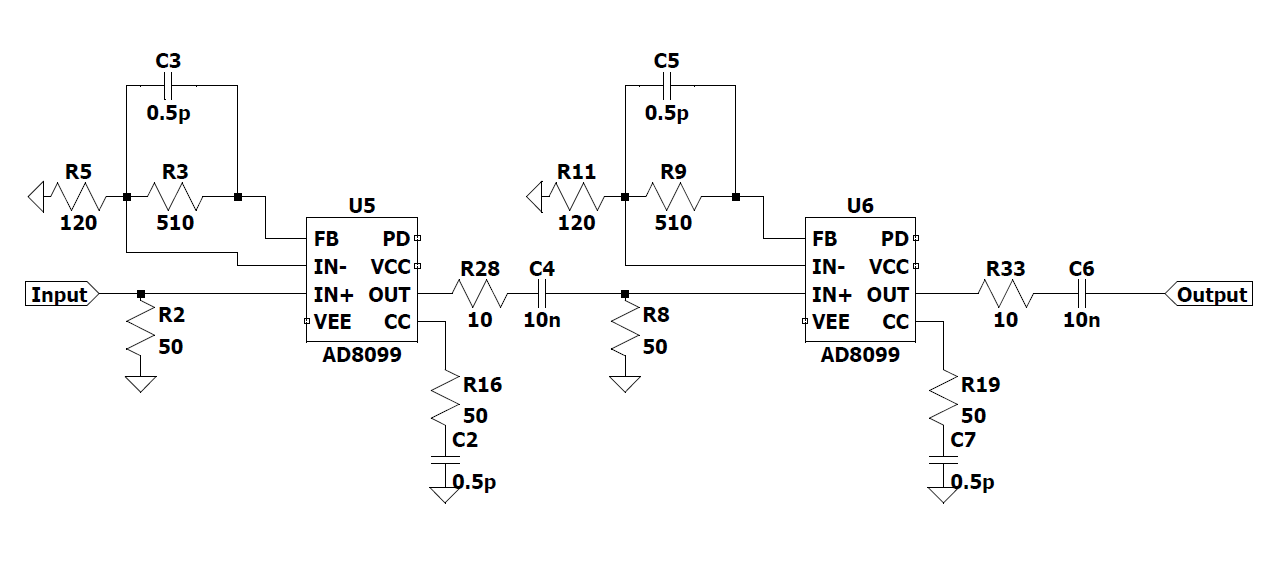}
 \caption{Schematic diagram of a two-cascade wide-bandwidth amplifier.}
\label{fig:fig4} 
\end{figure}

A simplified schematics  of a two-cascade amplifier  is shown
in Fig.\ref{fig:fig4}. 
The discrimination threshold must be in the range of $\sim$0.5 mV of the MRPC signal to obtain high detection efficiency. Since fast comparators are typically stable with thresholds down to about 10 mV, the MRPC signal must be amplified by a factor of $\sim$20.  
The fast  AD8099   amplifier \cite{AD8099} which has  a unity gain bandwidth up to 3 GHz was selected for the first cascades.  It has an overall DC gain of 19 (being 50-$\Omega$ loaded) and rise time of 1.1~ns measured and confirmed with LTspise simulations \cite{ltspice}. The optimal delay of the CFD for this circuit is estimated as 0.55~ns.

\begin{figure}[h]
\begin{minipage}[t]{0.47\textwidth}
\centering
\includegraphics*[width=70mm,height=60mm]{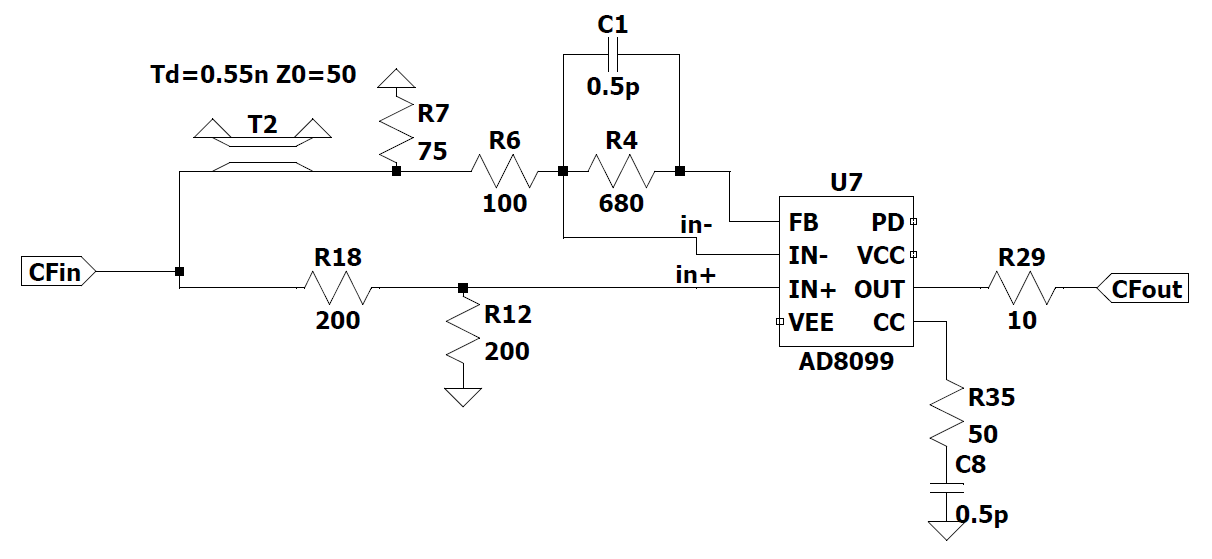}  
\end{minipage}\hfill
\begin{minipage}[t]{0.47\textwidth}
 \centering
\includegraphics*[width=70mm,height=60mm]{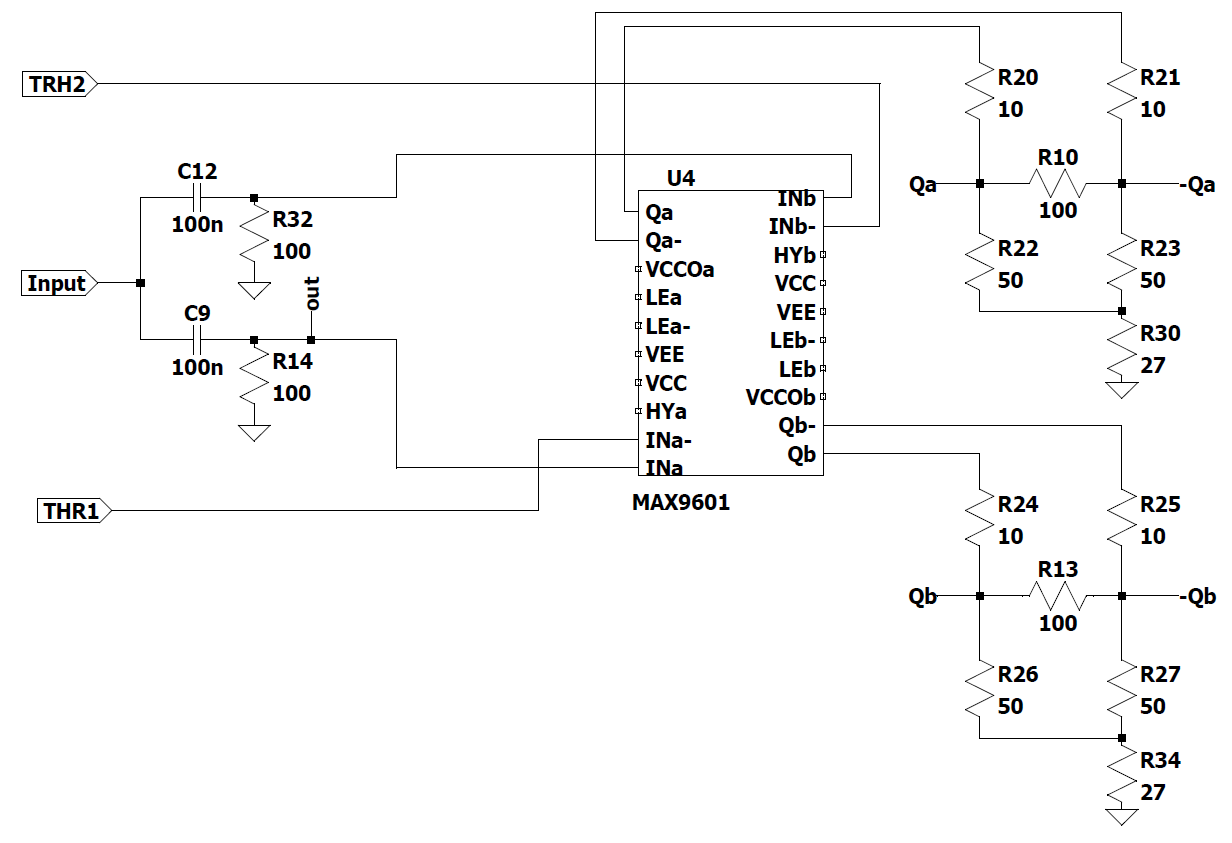} 
\end{minipage}
\caption{CFD circuit with delay/attenuation functions (left), the comparator cascade (right).  }
\label{fig:fig5}
\end{figure}

The CFD delay/attenuation cascade was also designed using the AD8099 amplifier\cite{AD8099}.
The delay was implemented with a piece of 50-$\Omega$ coaxial cable. It will be replaced by a 50-$\Omega$  PCB trace in the final design. The comparator cascade was built using an ultra-high speed MAX9601 comparator  \cite {MAX9601} which had two channels, used for low- and high thresholds. Both CFD and comparator cascades are shown in the left and right panels of Fig. \ref{fig:fig5}. 
The thresholds $V_1$ and $V_2$ were set to $\sim$0.4 mV and $\sim$0.8 mV of the amplifier input (MRPC output) signal.

The intrinsic time resolution of the readout chain was measured with triangular input pulse shape (rise and fall edges of 500 ps) taken from a pulse generator. The signal was split by a passive high frequency divider and sent to different CFD amplifier board channels. The time difference between two channels was measured with TDC64VHLE time-to-digit converter (TDC) module   \cite{TDC_LHEP} based on  HPTDC ASIC running in 25-ps binning mode with an intrinsic resolution of $\sim$17~ps.    The signal amplitude after the divider was varied from 2.8 to 34 mV.

\begin{figure}[hbtp] 
\centering
\includegraphics*[width=140mm,height=70mm]{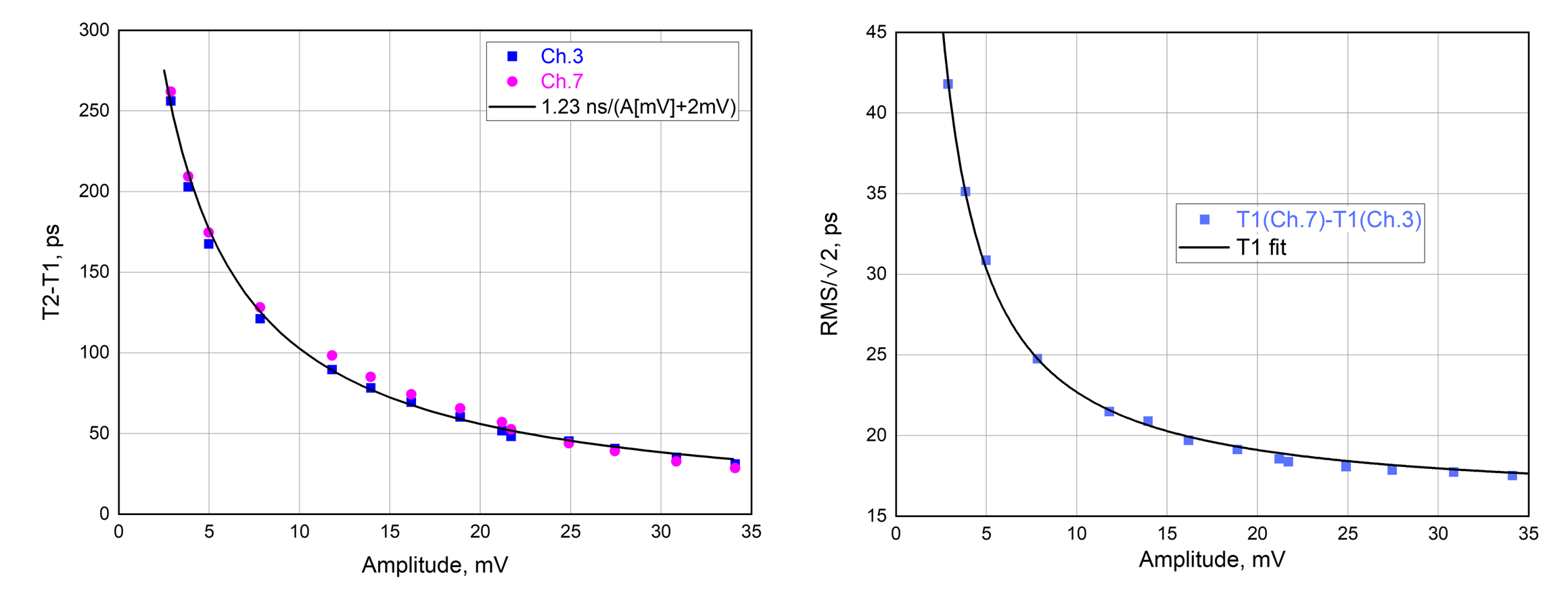}
\caption{Left: time difference $T_2$-$T_1$ for two TDC \cite{TDC_LHEP} channels shown by the solid symbols with the approximation by the inversed amplitude function given by the solid line. Right: time resolution for the $T_1$ signal, the line is the results of the approximation by the function (\ref{resolution}).}
\label{fig:fig6} 
\end{figure}

The time difference $T_2$-$T_1$ for two  TDC \cite{TDC_LHEP} channels as a function of the generator signal amplitude  is shown in the left panel of Fig.\ref{fig:fig6}. 
Measured points  shown by the solid squares and circles were approximated by the inversed amplitude function (solid line) as expected for the linear rising edge. 
However, it was found that an offset of 1.9 mV must be introduced for better approximation.
Some  offset is expected for the  MAX9601 comparator circuit  \cite {MAX9601}.
The time resolution as {a function of the generator signal amplitude} was estimated from the time difference { $T_1$(Ch.7)-$T_1$(Ch.3) for two 
different FEE channels using two  TDC \cite{TDC_LHEP} channels.}  
The resulting  time resolution was obtained dividing the RMS value by $\sqrt{2}$ (assuming TDC channels were identical). The result is shown in the right panel  of Fig.\ref{fig:fig6}  together with the best fit of the data by the resolution function taken in the following form 
\begin{eqnarray}
\label{resolution}
RMS =C_0 \bigoplus \frac{C_1}{\sqrt{A}}  \bigoplus \frac{C_2}{A},
\end{eqnarray}
where $A$ is the signal amplitude in mV, $C_0$, $C_1$  and $C_2$ are the constant,
jitter and noise term contributions, respectively. 
The constant term $C_0$ is consistent with the expected TDC resolution of $\sim$17~ps.
The intrinsic resolution is dominated by the TDC contribution $C_0$ for the signal amplitudes above 10 mV. The resolution for signals with smaller amplitudes is dominated by the 
electronics jitter $C_1$ and  noise $C_2$ terms. 

\section{MRPC test with cosmic muons}
\label{cosmics}

\begin{figure}[hbtp]
\begin{minipage}[t]{0.47\textwidth}
\centering
\includegraphics*[width=60mm,height=60mm]{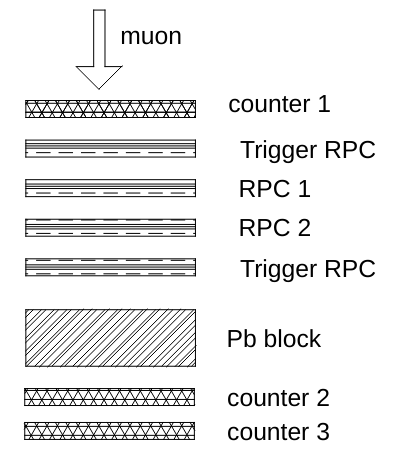}  
\end{minipage}\hfill
\begin{minipage}[t]{0.47\textwidth}
 \centering
\includegraphics*[width=60mm,height=60mm]{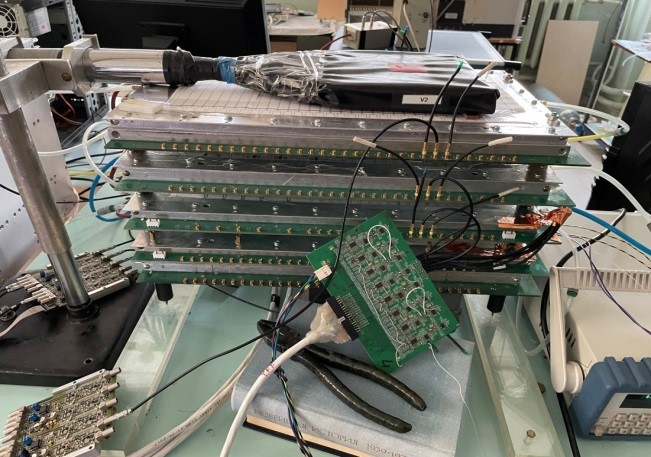} 
\end{minipage}
\caption{Scheme and picture of the cosmic test setup.  }
\label{fig:fig7}
\end{figure}

The readout chain was tested and optimized with detecting cosmic particles. The setup is shown in Fig. \ref{fig:fig7}. It consists of  four MRPCs, three scintillator counters and a block of lead to filter out low-energy particles. The two inner MRPCs were used for the signal  measurements, while the two outer chambers were included in the trigger logics similar to   the MRPC signal shape  measurements described in   Section~\ref{MRPC}.
The distance between the MRPCs was 48 mm. The chambers were aligned horizontally with an accuracy of $\pm$0.3 mm. The scintillation counters had 18$\times$18~cm$^2$ active area covering the full length of the chamber strips. The lead block had the thickness of 9~cm absorbing muons with a kinetic energy below $\sim$200 MeV ($\beta$>0.94). Therefore,  the time of flight between the two studied MRPCs varied as low as $\sim$10 ps 
for these particles providing negligible contribution to the total time resolution. 
The trigger was formed as a coincidence of signals from the three scintillation counters and 1 strip of each of the trigger MRPCs.   The trigger rate was about 0.05 Hz.  Measurements were performed with a full readout chain described in Section~ \ref{design}.

\begin{figure}[hbtp] 
\centering
\includegraphics*[width=80mm,height=80mm]{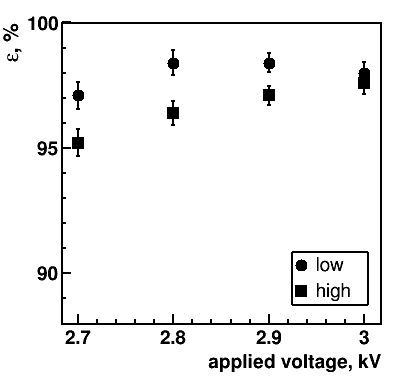}
\caption{Efficiency of the 12-gap MRPC. Solid circles and squares are the results for low and high thresholds, respectively.}
\label{fig:fig8} 
\end{figure}
 
The detection efficiency $\epsilon$ of the 12-gap MRPC as a function of the applied high voltage is shown 
in Fig.\ref{fig:fig8}. The CFD thresholds were set to $V_1$ = 0.45 mV and $V_2$ = 0.85 mV.
The efficiency was defined as a ratio of the numbers of the signals with the  
amplitudes above the low threshold $V_1$ (solid circles) or high threshold $V_2$ (solid squares) 
to the  number of triggers. It was found $\sim$96\% for chamber bias above 2.8 kV.

\begin{figure}[hbtp]
\begin{minipage}[t]{0.47\textwidth}
\centering
\includegraphics*[width=70mm,height=70mm]{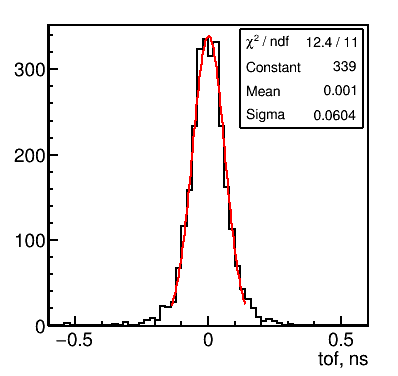}  
\caption{MRPC $T_0$ time resolution.}
\label{fig:fig9}
\end{minipage}\hfill
\begin{minipage}[t]{0.47\textwidth}
 \centering
\includegraphics*[width=70mm,height=70mm]{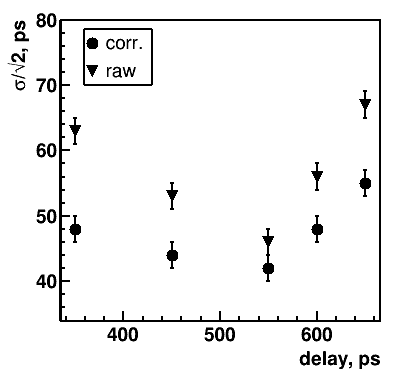} 
\caption{CFD delay optimization.  }
\label{fig:fig10}
\end{minipage} 
\end{figure}

The time resolution was measured with HV bias of 2.85~kV with a CFD fraction fixed to 0.5. The time reference $T_0$ was calculated for each MRPC according to an iterative algorithm described in 
\cite{kuzmin}. 
The time difference $T_0$ for the two MRPCs under test is presented   in Fig.\ref{fig:fig9}. The line is the result of the approximation by a gaussian function. 
The time resolution of the stand-alone MRPC was obtained by dividing   the obtained standard deviation from the gaussian fit by  $\sqrt{2}$  (assuming both MRPCs contribute identically).

The time resolution of a single MRPC is shown in Fig.\ref{fig:fig10} as a function of the CFD delay varied from 0.45 ns to 0.65 ns.  The solid triangles and circles represent the results for the time differences of the
low threshold comparator $T_1$ and zero crossing $T_0$, respectively.
Best $T_0$ time resolution of 42 ps was achieved with a 0.55 ns delay, as expected from the MRPC pulse shape measurements. 
Note that the difference between the low threshold ($T_1$) and  zero crossing ($T_0$) time resolutions was  negligible at the optimal CFD delay of 0.55~ns.  

\section{Muon beam test}
\label{beamtest}

\begin{figure}[hbtp] 
\centering
\includegraphics*[width=80mm,height=80mm]{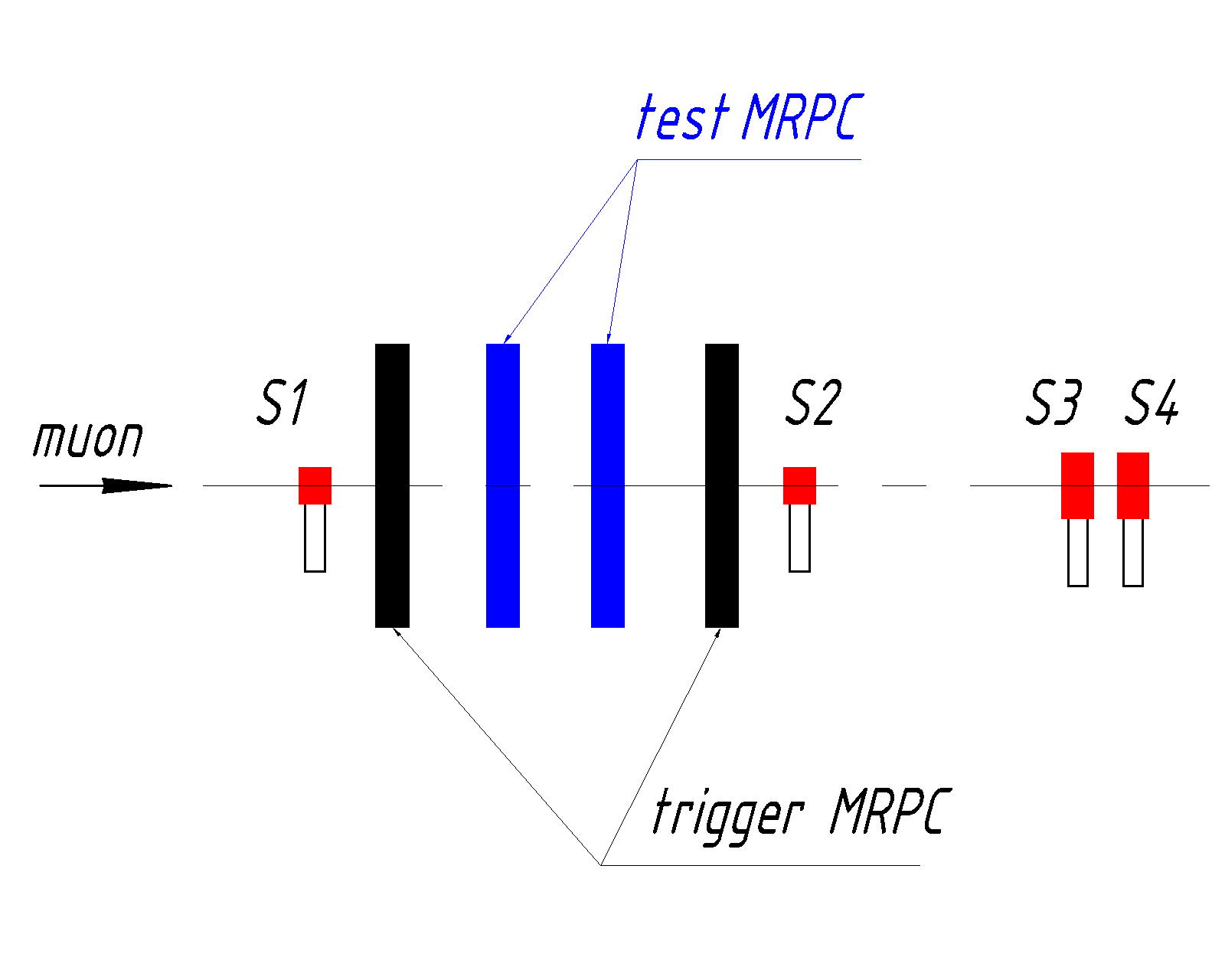}
\caption{Schematic view of the MRPC test setup. $S1$-$S4$ are the scintillation counters, 
the test and trigger MRPCs are the chambers under test and included in the trigger.}
\label{fig:fig11} 
\end{figure}

The MRPC test setup is schematically shown in Fig. \ref{fig:fig11}. The data were
taken using a muon beam at the U-70 accelerator in Protvino \cite{U70}. The
muons were originated from the interaction of the circulating proton
beam with the internal target placed in the accelerator vacuum chamber.
The duration of the beam spill was about  0.3~s with a repetition rate of 0.1 Hz.
No momentum selection was applied to the beam. The averaged momentum of muons was about 2 GeV/$c$. The beam intensity was 100$\div$10k muons/s/cm$^2$ in the test area. 
The setup consists of four scintillation counters ($S1$-$S4$), two trigger MRPCs and two MRPCs under test. Small-size counters $S1$, $S2$ formed an active beam area of 1$\times$1 cm$^2$.
The signals from the trigger MRPCs were included in trigger logics, as described in previous sections. They also were used to align the detectors on the beam axis.
The low and high CFD thresholds were increased up to $V_1$ = 0.8 mV and $V_2$ = 1.4 mV, respectively, to reduce possible noise contribution. The CFD delay was set initially to 0.55 ns according to the results of the measurements with cosmic muons. The MRPC efficiency was measured  to be 99\% at HV bias above 2.75 kV.
 
\begin{figure}[hbtp] 
\centering
\includegraphics*[width=140mm,height=70mm]{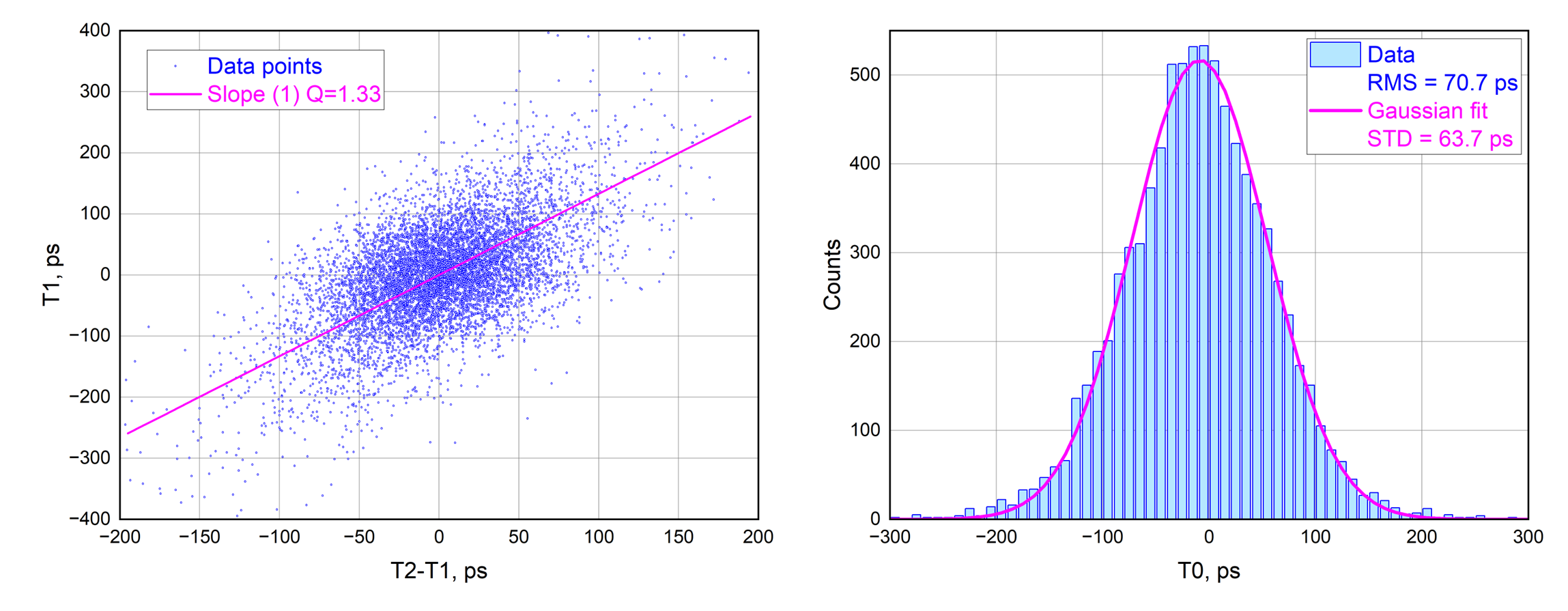}
\caption{Left: correlation of $T_1$ versus $T_2$-$T_1$, right: $T_0$ time difference for the MRPCs under test calculated according to (\ref{t0}).}
\label{fig:fig12} 
\end{figure}

The correlation of the $T_1$ and $T_2-T_1$ differences for the two MRPCs under test 
is demonstrated in the left panel of Fig.\ref{fig:fig12}. The line is the result of simple linear correction according to (\ref{t0}) with the factor $Q = 1/(R-1)$ = 1.33 (for set thresholds).
One can see good agreement of the $T_1$ and $T_2-T_1$ data correlation with the linear approximation (\ref{t0}). 
The $T_0$ distribution  for the two MRPCs   is shown in the right panel of Fig.\ref{fig:fig12} together with  the gaussian approximation result.  The RMS from the data and standard deviation from the gaussian fit are $\sim$71~ps and $\sim$64~ps, respectively.
 
\begin{figure}[hbtp] 
\centering
\includegraphics*[width=140mm,height=70mm]{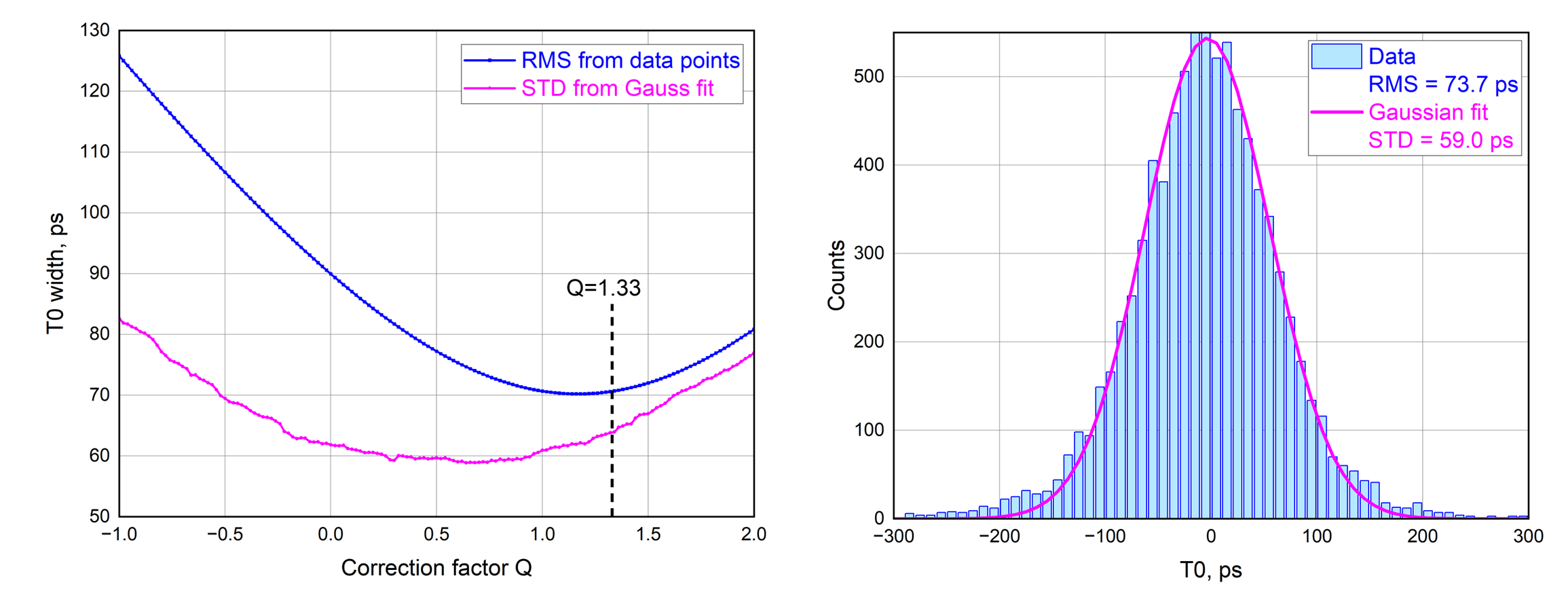}
\caption{Left: Optimization of the correction factor $Q$, right: $T_0$ time difference  
at $Q$=0.7.}
\label{fig:fig13} 
\end{figure}

Linear correction for $T_0$ according to (\ref{t0}) is valid only in the  absence of electronics noise.  However, its contribution  is non negligible as one can see from the left panel of 
Fig.\ref{fig:fig6}. 
The $T_0$ distribution was analyzed for different values of the correction factor $Q$    varied in the range from -1 up to +2. 
The values  $Q$=-1 and $Q$=0 corresponded to   $T_0$=$T_2$ and $T_0$=$T_1$,
respectively. The result is presented in the left panel of Fig.\ref{fig:fig13}.
Both minimal values of the RMS from the data and standard deviation from the fit are reached
at $Q\approx$1.16 and $Q\approx$0.7, respectively. The $T_0$ distribution for the correction factor $Q$ = 0.7 is shown in the right panel of Fig.\ref{fig:fig13}.  
In this case the RMS from the data and standard deviation from the fit are $\sim$74~ps and $\sim$59~ps, respectively.
 
\begin{figure}[hbtp] 
\centering
\includegraphics*[width=140mm,height=70mm]{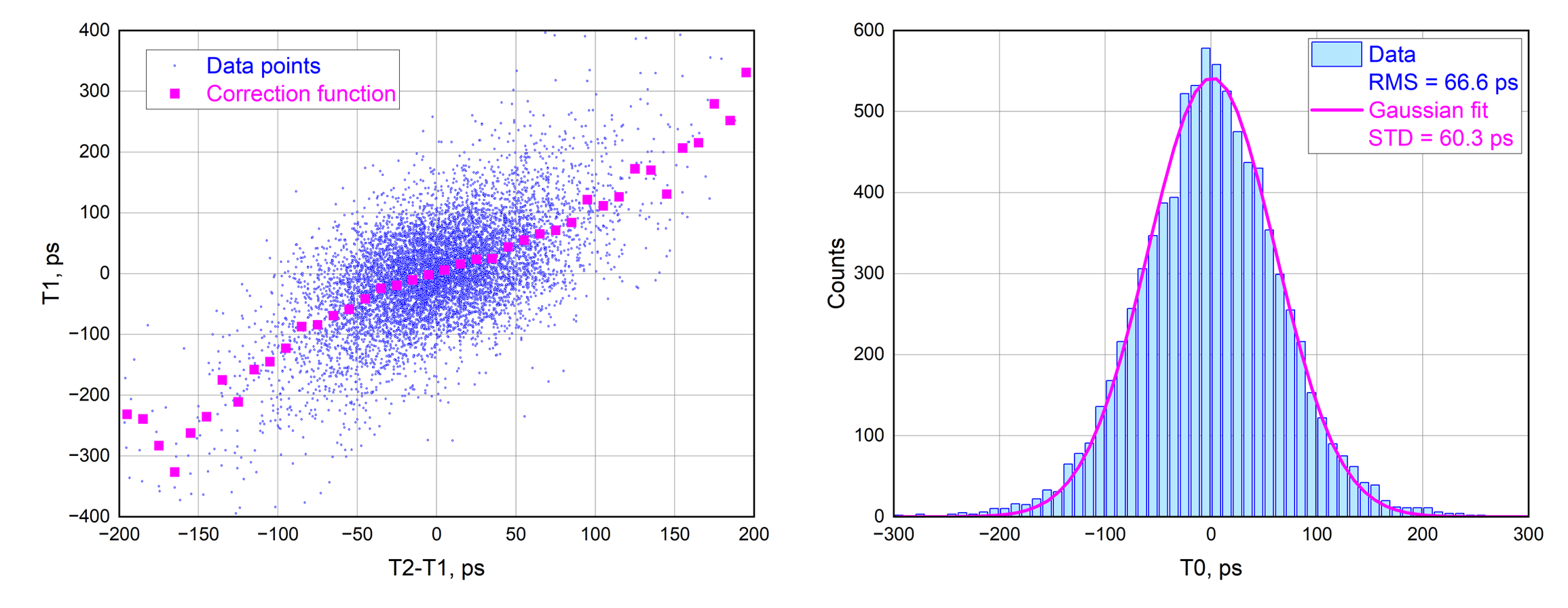}
\caption{Left: correlation of $T_1$ versus $T_2$-$T_1$ with the correction function $C(T_2 -T_1)$, right: $T_0$ time difference for the MRPCs under test calculated according (\ref{t0c}). }
\label{fig:fig14} 
\end{figure}

In general, signal rise is nonlinear (see Fig.\ref{fig:fig2}), thus the $T_2$-$T_1$  difference can be nonlinear. In this case the expression for the $T_0$ value (\ref{t0}) can be rewritten in a more general form:
\begin{eqnarray}
\label{t0c}
T_0 = T_1 - C(T_2 -T_1),  
\end{eqnarray}
where $C(T_2 -T_1)$ is a correction  function that minimizes the $T_0$ dispersion (either RMS from the data or standard deviation from the fit) for the given data set.
This function can be found with the iterative algorithm described in Ref.\cite{kuzmin}.
The $C(T_2 -T_1)$ function for the data set used for Fig.\ref{fig:fig12} is shown by the solid points in left panel of Fig.\ref{fig:fig14}.
One can see that $C(T_2 -T_1)$ is almost linear for the $T_1$-($T_2$-$T_1$) correlation center while some non-linearity appears only for the distribution tails.  $C(T_2 -T_1)$ was
parameterized by polynomials  for further practical use.
The resulting $T_0$ distribution  for the two MRPCs  calculated using expression (\ref{t0c})  is shown in the right panel of Fig.\ref{fig:fig14} together with  the gaussian approximation result. The RMS from the data and standard deviation from the fit are $\sim$67~ps and $\sim$60~ps, respectively.
The standard deviation value is almost the same as in the case of optimized linear correction 
(see right panel of Fig.\ref{fig:fig13}) because $C(T_2 -T_1)$ is practically linear 
in the area of main peak. The RMS value for the data is smaller than in the case of optimized linear correction.  Therefore, the data obtained with the muon beam were analyzed using a  nonlinear algorithm.

\begin{figure}[hbtp]
\begin{minipage}[t]{0.47\textwidth}
\centering
\includegraphics*[width=70mm,height=70mm]{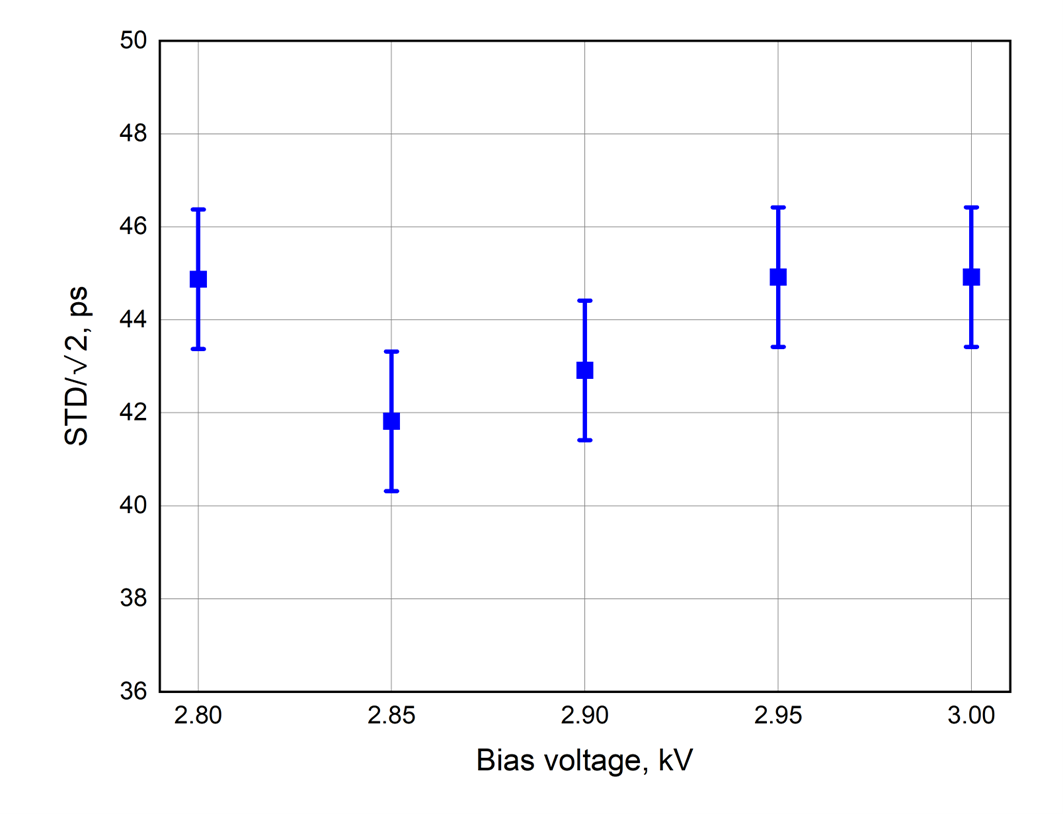}  
\caption{MRPC   time resolution as a function of HV setting.}
\label{fig:fig15}
\end{minipage}\hfill
\begin{minipage}[t]{0.47\textwidth}
 \centering
\includegraphics*[width=70mm,height=70mm]{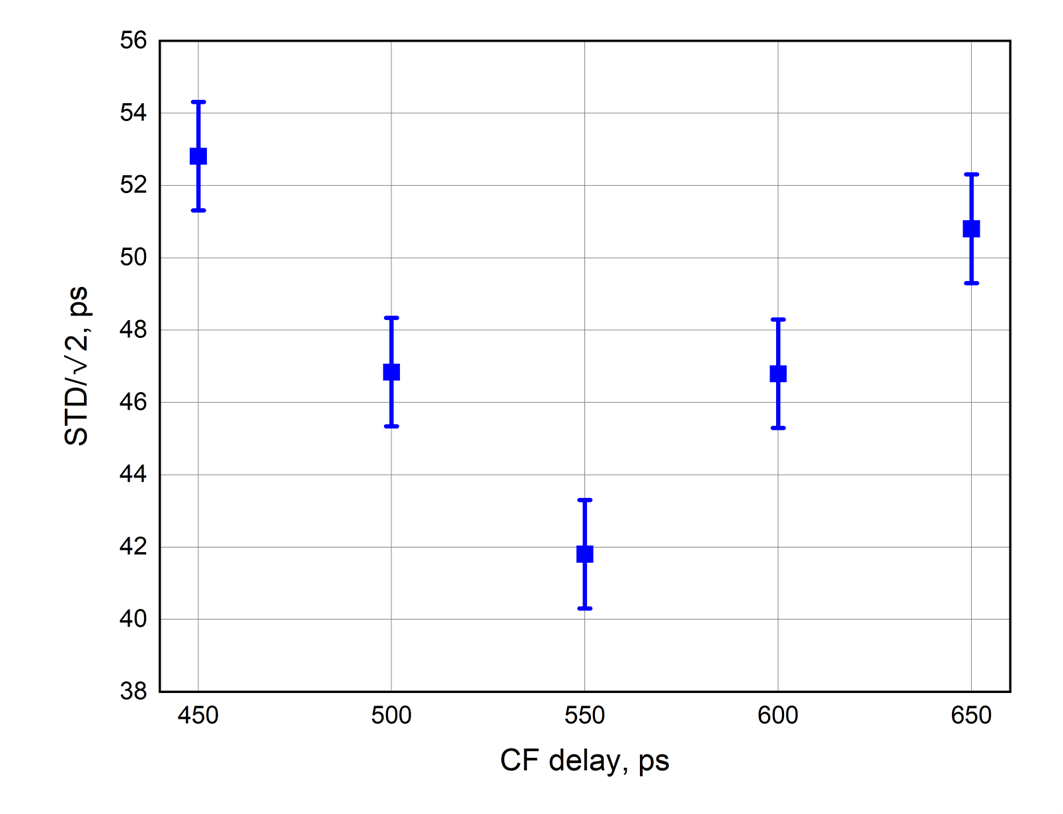} 
\caption{MRPC   time resolution as a function of CFD delay.  }
\label{fig:fig16}
\end{minipage} 
\end{figure}

The resulting time resolution of a single MRPC was estimated by dividing   the $T_0$ standard deviation by  $\sqrt{2}$, assuming the two chambers contribute identically. Fig.\ref{fig:fig15} shows the time resolution when bias of one chamber was fixed while that of the second chamber was varied. It was found that the optimal HV bias is 2.85~kV for both chambers. However, 
only weak HV dependence of the time resolution was observed over the full MRPCs working range. 

The effect of CFD delay on time resolution is demonstrated in Fig.\ref{fig:fig16}. This measurement was made with a fixed 0.55~ns delay for the 10-gap MRPC CFD by varying  the delay for the 12-gap chamber CFD. As expected, it was found that the optimal CFD delay for both chambers is 0.55~ns as already determined in the cosmic test.

\section{Conclusions}

The CFD discriminator for MRPC readout is designed and implemented.  
the intrinsic time resolution of the readout chain measured with a waveform generator is 
15-30 ps for input amplitudes of 3-30 mV.  
 
The CFD concept is validated for  10- and 12- gap MRPCs with cosmic muons and with a 2 GeV/$c$ muon beam. 
The MRPC total time resolution achieved with optimized CFD parameters (thresholds and delay time) is  $\sim$40~ps being consistent with the results obtained using time reconstruction methods based on   neural networks and ToT correction \cite{WF1}.

Further improvement of the CFD discriminator for MRPC readout can be related to use of
faster amplifiers with lower power consumption. In future, the  CFD concept 
can be implemented in a new front-end ASIC  for MRPC  similar to VFAT3 ASIC 
\cite{VFAT3} developed for GEM detectors. 
 







\end{document}